\documentclass[aps,prl,twocolumn,footinbib,showpacs,longbibliography]{revtex4-1}
\usepackage{amsmath}
\usepackage{amssymb}
\usepackage{hyperref}
\usepackage{color}
\usepackage{natbib}

\usepackage{bm}
\usepackage{epsfig}

\newcommand{\rmi}{{\rm i}}
\newcommand{\e}{{\rm e}}

\hypersetup{pdfstartview={FitH},pdfpagemode={UseNone},
            colorlinks,linkcolor=red, citecolor=blue, urlcolor=blue,
            bookmarks=true, bookmarksopen=true, pdfnewwindow=true}

\begin{document}

\title{Biexciton-mediated superradiant photon blockade}

\author{\firstname{Alexander~V.} \surname{Poshakinskiy}}
\email{poshakinskiy@mail.ioffe.ru}
\affiliation{Ioffe  Institute, St.~Petersburg 194021, Russia}
\author{\firstname{Alexander~N.} \surname{Poddubny}}
\affiliation{Ioffe  Institute, St.~Petersburg 194021, Russia}



\begin{abstract}
The photon blockade is  a hallmark of quantum light transport through a single  two-level system that  can accomodate only one photon. Here, we theoretically show that two-photon transmission can be suppressed  even  for a seemingly classical system with large number of quantum dots in a cavity when the biexciton nonlinearity is taken into account.
We reveal the nonmonotonous dependence of the  second-order correlation function of the transmitted photons on the biexciton binding energy. The blockade is realized by proper tuning the biexciton resonance that controls the  collective superradiant modes.
\end{abstract}

 \maketitle

{\it Introduction.}---A two-level system can absorb or emit only one photon at a time~\cite{Carmichael,Diedrich1987}. As such, it is an ultimate quantum playground  for the photon blockade realization~\cite{Imamoglu1997}. Antibunching of transmitted photons in the photon blockade regime has been recently demonstrated for a cavity strongly coupled to a single atom~\cite{Kimble2005,Kubanek2008,Schuster2008} and to a single quantum dot~\cite{Ates2009,Faraon2008,Hu2012,Pan2013,Vucovic2015}. Increasing the number of dots, one brings the system towards the classical limit. In particular, after the first photon has been absorbed, the second one can be still accommodated in one of the unoccupied dots and the photon blockade is destroyed. Indeed, the previous theoretical results for atoms without a cavity yield the absence of photon blockade~\cite{Yudson1984,Yudson2008}.  While the photon bunching for ensemble of two-level atoms strongly coupled to a cavity has been predicted in Ref.~\cite{Brecha1999}, the fundamental question of photon blockade feasibility for multiple quantum emitters remains open. 
Existing studies of photon blockade in cavities were limited to the case of single resonant atom~\cite{Shi2011,Shi2013}.
Here, we show that contrary to the na\"ive expectations, the photon blockade effect can be realized in a cavity with an arbitrary number of quantum dots provided that their biexcitonic nonlinearity is exploited. The reason is that the two-photon transport probes only collective superradiant modes controlled by biexciton quantum dot resonances rather than all quantum dot modes. Therefore the photon blockade is attained when  the superradiant modes are detuned from the two-photon resonance.

%

\begin{figure}[b]
 \includegraphics[width=.7\columnwidth]{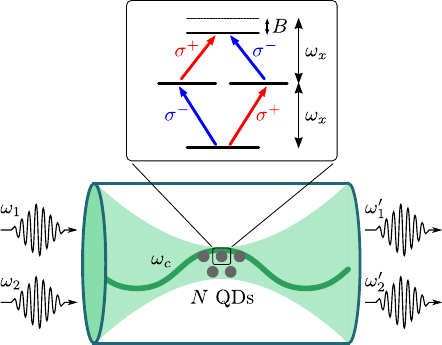}
 \caption{A sketch of the two-photon transmission through a cavity with $N$ QDs. The inset illustrates the level scheme of a QD with the biexciton nonlinearity. }\label{fig:pic}
\end{figure}

{\it Model.}---We consider two-photon transport through a microcavity with $N$ embedded quantum dots (QDs), see Fig.~\ref{fig:pic}. The Hamiltonian describing the ensemble of QDs coupled to the cavity mode reads
\begin{align}\label{eq:H}
H = &\sum_{\sigma=\pm} \Big\{ \omega_c c^\dag_\sigma c_\sigma^{\vphantom{\dag}} + \sum_{i=1}^N \left[ \omega_x b^\dag_{i,\sigma} b_{i,\sigma}^{\vphantom{\dag}}   + g (b_{i,\sigma}^\dag c_\sigma^{\vphantom{\dag}} + c_\sigma^\dag b_{i,\sigma}^{\vphantom{\dag}}) \right] \Big\} \nonumber\\
&-B \sum_{i=1}^N b^\dag_{i,-} b^\dag_{i,+} b_{i,-}^{\vphantom{\dag}} b_{i,+}^{\vphantom{\dag}} \,,
\end{align}
where $c_\pm$ are the annihilation operators corresponding to the two degenerate circularly polarized cavity modes with the frequency $\omega_c$, $b_{i,\pm}$  is the lowering operator of the two-level system corresponding to annihilation of an exciton with spin projection $\pm 1$ in the $i$-th QD, $\omega_x$ is the exciton resonance frequency, $g$ is the exciton-photon coupling strength, and $B$ is the biexciton binding energy~\cite{Benson2000,Langbein2013}. The dagger denotes Hermitian conjugation and $\hbar =1$. We suppose here that all QDs are identical and neglect the polarization splitting of the cavity mode and exciton for simplicity.

{\it Method.}---Elaborate  theoretical methods including Bethe ansatz~\cite{Yudson1984, Yudson2008,Fan2007}, Lippman-Schwinger equation~\cite{Baranger2013}, input-output formalism~\cite{Fan2010}, Green's functions~\cite{Busch2015}, and path integral~\cite{Shi2009,Ringel2013} have been previously used to study
two-photon transport in quantum systems, such as atoms in a cavity~\cite{Shi2011,Shi2013} or in a waveguide~\cite{Laakso2014,Rephaeli2011,Fang2015}.
However, their straightforward application to multiple QDs with biexciton nonlinearity in a cavity appears to be complicated.
The intuitive representation of the two-photon scattering process given by Feynman diagram technique~\cite{LL4} allows one to obtain the scattering amplitude only in the lowest orders of perturbation theory. 
Here, in order to exactly sum the perturbation series, 
we regard the operators $b_{i,\sigma}$ as bosonic.  Correspondence with the problem where the operators $b_{i,\sigma}$ describe two-level systems is established by adding to the Hamiltonian an auxiliary interaction term
$
 \sum_{\sigma=\pm} \sum_{i=1}^N Vb^\dag_{i,\sigma} b^\dag_{i,\sigma} b_{i,\sigma} b_{i,\sigma} \,
$
and considering the limit $V \to \infty$~\cite{Busch2010,Roy2010,Baranger2013}. In this limit the interaction with unphysical states 
that have more than one excitation $b_{i,\sigma}^\dag$
is absent due to their infinite energy separation from the physical states. Surprisingly, this procedure drastically simplifies the diagrammatic approach. Moreover, it naturally handles the biexciton nonlinearity in Eq.~\eqref{eq:H} that has the same form as the auxiliary interaction term. 

\begin{figure}
 \includegraphics[width=\columnwidth]{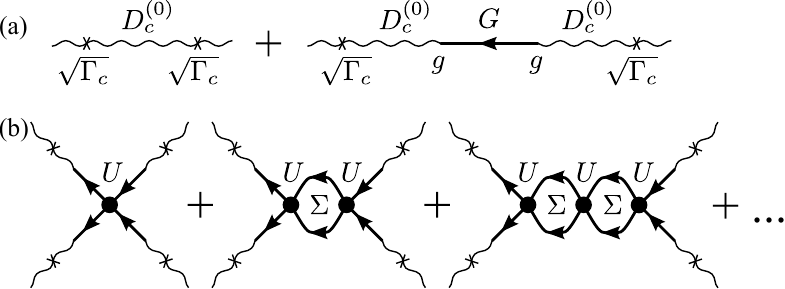}
 \caption{ 
 (a) Diagrammatic representation of the single-photon transmission. Wavy line is the bare cavity mode Green's function, solid straight line is the dressed exciton Green's function Eq.~\eqref{eq:G}, the crosses indicate photon entering/exiting the cavity. 
 (b) The series corresponding to the two-photon scattering. The dots represent interaction of two excitons.}\label{fig:dia}
\end{figure}

We start by calculating Green's function $G_{ij}(\omega)$ describing propagation of an exciton from $j$-th to $i$-th QD. 
Green's function is polarization-independent so we omit the corresponding indices. The Dyson equation describing the dressing of exciton by the cavity mode reads~\cite{Shi2011},  
$G_{ij} = \delta_{ij} G^{(0)} + N g^2   G^{(0)} D_c^{(0)} \sum_{k=1}^N G_{kj}$, 
where $G^{(0)}(\omega)=1/(\omega-\tilde\omega_x)$ is bare exciton Green's function and $D_c^{(0)} = 1/(\omega-\tilde\omega_c)$ is bare cavity mode Green's function. We introduced here $\tilde\omega_c = \omega_c - \rmi\Gamma_c$ and $\tilde\omega_x = \omega_x - \rmi\Gamma_x$, with $\Gamma_c$ being the decay rate of the cavity mode due to light escape through the mirrors and $\Gamma_x$ being the exciton nonradiative decay rate.
Solution of Dyson equation gives the dressed exciton Green's function
\begin{align}\label{eq:G}
G_{ij}(\omega) = \frac{1}{\omega-\tilde\omega_x}\left[ \delta_{ij} + \frac{g^2}{(\omega-\tilde\omega_x)(\omega-\tilde\omega_c)-Ng^2} \right] \hspace{-.1cm}. 
\end{align}
Using Green's function Eq.~\eqref{eq:G} we can find the transmission coefficient for a single photon. Summing two contributions to this process shown in Fig.~\ref{fig:dia}(a) we get~\cite{kavbamalas}
\begin{align}\label{eq:t}
t(\omega) = -\frac{\rmi\Gamma_c(\omega-\tilde\omega_x)}{(\omega-\tilde\omega_x)(\omega-\tilde\omega_c)-Ng^2} \,.
\end{align}
We used here that the amplitude of the photon entering and leaving the cavity (crosses in Fig.~\ref{fig:dia}) is equal to $\sqrt{\Gamma_c}$.

We are now in position to consider two-photon transmission through the system. The process $(\omega_1\sigma_1,\omega_2\sigma_2)\to(\omega_1'\sigma_1',\omega_2'\sigma_2')$ is described by the scattering matrix that can be written in the form~\cite{Xu2013} 
\begin{align}\label{eq:S}
S_{1'2',12} = (2\pi)^2
  &[\delta_{\sigma_1'\sigma_1}\delta_{\sigma_2'\sigma_2}\delta(\omega_1'-\omega_1)\delta(\omega_2'-\omega_2)\nonumber \\
+&\delta_{\sigma_2'\sigma_1}\delta_{\sigma_1'\sigma_2}\delta(\omega_2'-\omega_1)\delta(\omega_1'-\omega_2)]  t(\omega_1)t(\omega_2) \nonumber\\
+&2\pi\rmi M_{1'2',12}\,  \delta(\omega_1'+\omega_2'-\omega_1-\omega_2) \,.
\end{align}
The first two terms in Eq.~\eqref{eq:S} correspond to independent transmission of the photons without a change of their frequencies. Such  processes are described by the single-photon transmission coefficient Eq.~\eqref{eq:t}.
The last term in Eq.~\eqref{eq:S} describes the process of photons scattering on each other when their frequencies do change. 
The two-photon scattering occurs as follows, see Fig.~\ref{fig:dia}(b): First, the photons enter the cavity and turn into excitons. Then the excitons interact with each other due to infinite repulsion potential (for excitons with parallel spins) or the finite biexciton binding energy $B$ (for excitons with opposite spins). Finally, the excitons turn back into photons and exit the cavity. 
The exact scattering amplitude is given by the series shown in Fig.~\ref{fig:dia}(b). Summation of the series yields 
\begin{align}\label{eq:M}
\hspace{-.2cm} M_{1'2',12} &= \sum_{i,j=1}^N \left[ U (1+\Sigma U)^{-1} \right]_{i,\sigma_1'\sigma_2';j,\sigma_1\sigma_2} \prod_a s(\omega_a), 
\end{align}
where 
$
s(\omega_a)=g\sqrt{\Gamma_c}/[(\omega_a-\tilde\omega_x)(\omega_a-\tilde\omega_c)-Ng^2] 
$
describes the outer ends of the diagrams with $\omega_a=\omega_1,\omega_2,\omega_1',\omega_2'$ being the frequencies of the incoming and outgoing photons, matrix
\begin{align}\label{eq:Sigma}
\hspace{-.2cm}\Sigma_{i,\sigma_1'\sigma_2';j,\sigma_1\sigma_2} = \rmi \delta_{\sigma_1\sigma_1'}\delta_{\sigma_2\sigma_2'} \int  G_{ij}(\omega)G_{ij}(2\varepsilon-\omega) \frac{d\omega}{2\pi} 
\end{align}
with $\varepsilon=(\omega_1+\omega_2)/2=(\omega_1'+\omega_2')/2$ being the mean photon energy describes propagation of two excitons with spins $\sigma_1$ and $\sigma_2$ from $j$-th to $i$-th QD, and $U_{i,\sigma_1'\sigma_2';j,\sigma_1\sigma_2}=\delta_{ij}U_{\sigma_1'\sigma_2',\sigma_1\sigma_2}$ with $U_{++,++}=U_{--,--} \to \infty$, $U_{+-,+-}=U_{+-,-+}=U_{-+,+-}=U_{-+,-+}=B/2$ describes the  interaction between excitons.

%
%

{\it Two-photon resonances.}---Substituting Green's function Eq.~\eqref{eq:G} into Eqs.~\eqref{eq:M}--\eqref{eq:Sigma} we obtain the scattering amplitudes for photons with parallel and opposite spins,
\begin{align}\label{eq:Mpp}
&M_{\uparrow\uparrow}\equiv M_{++,++}= M_{--,--}
\\ \nonumber
&=\frac{4N(2\varepsilon-\tilde\omega_x-\tilde\omega_c)(\varepsilon-\tilde\omega_x-\frac{Ng^2}{\varepsilon-\tilde\omega_c}) \prod_a s(\omega_a)}{2\varepsilon-\tilde\omega_x-\tilde\omega_c-\frac{Ng^2}{\varepsilon-\tilde\omega_c}-\frac{(N-1)g^2}{\varepsilon-\tilde\omega_x}} ,
\\
\label{eq:Mpm}
&M_{\uparrow\downarrow}\equiv M_{+-,+-}= M_{+-,-+}=M_{-+,+-}= M_{-+,-+}
\\ \nonumber
&\hspace{-.3cm}=\hspace{-.1cm} \frac{2NB(2\varepsilon-\tilde\omega_x-\tilde\omega_c)(\varepsilon-\tilde\omega_x-\frac{Ng^2}{\varepsilon-\tilde\omega_c}) \prod_a s(\omega_a)}{(2\varepsilon-2\tilde\omega_x+B)[2\varepsilon-\tilde\omega_x-\tilde\omega_c-\frac{Ng^2}{\varepsilon-\tilde\omega_c}-\frac{(N-1)g^2}{\varepsilon-\tilde\omega_x}]-2g^2}.
\end{align} 
Two photons with the same spin do not excite the biexciton, so the amplitude $M_{\uparrow\uparrow}$ does not depend on $B$. Conversely, scattering of the photons with opposite spins is only due to non-zero exciton binding energy, so $M_{\uparrow\downarrow}=0$ at $B=0$. In the limit of large binding energy $B\to\infty$ interaction with biexciton state is suppressed so a $\sigma^+$ photon and a $\sigma^-$ one interact with the system exactly in the same way as two $\sigma^+$ photons. Thus, we have $M_{\uparrow\downarrow}=M_{\uparrow\uparrow}/2$, where the factor $2$ is due to indistinguishability of photons with parallel spins. 

\begin{figure}[t]
 \includegraphics[width=.99\columnwidth]{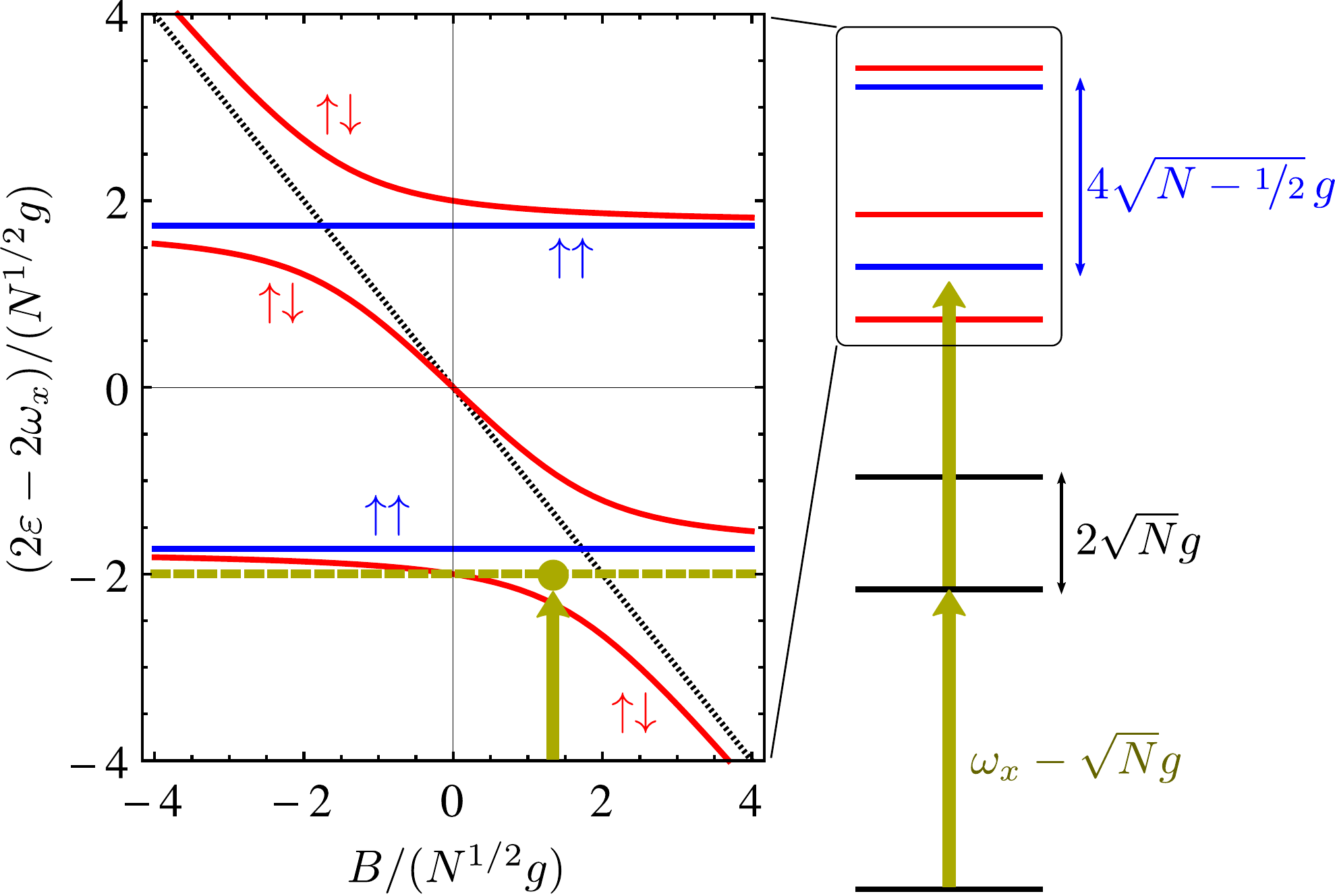}
 \caption{The energies of the two-photon resonances of the scattering amplitude for a cavity with $N=2$ QDs as functions of the biexciton binding energy $B$. The solid blue (red) curves show the states relevant for excitation by two photons with parallel (opposite) spins and correspond to the poles of Eq.~\eqref{eq:Mpp} [Eq.~\eqref{eq:Mpm}]. The black dotted line is the energy of the bare biexciton state. The yellow dashed line shows the energy of excitation, corresponding to the single-photon resonance. The dot indicates the optimal binding energy. Right panel sketches the photon blockade effect.}\label{fig:g2B}\label{fig:ladder}
\end{figure}

The poles of the scattering amplitude correspond to the eigenstates of the system. We start  from 
analysis of the one-photon resonances, being the poles of the factors $s(\omega_a)$ in the amplitudes Eq.~\eqref{eq:Mpp}--\eqref{eq:Mpm} and the transmission coefficient Eq.~\eqref{eq:t}. At $\omega_c=\omega_x$, $\Gamma_x=\Gamma_c=0$ the one-particle eigenfrequencies read $\omega_a = \omega_x \pm \sqrt Ng$, which corresponds to the energies of the first rung of the Tavis--Cummings ladder~\cite{Tavis1968}. The splitting $2\sqrt N g$ originates  from interaction of the symmetric superposition of excitons in all QDs (superradiant mode \cite{Dicke1954,Brecha1999,scully2006volga,Khitrova2007nat,Temnov2009,JETP2009}) with a cavity photon. In addition to these one-particle resonances, the scattering amplitude has also poles corresponding to two-particle excitations. For the amplitude $M_{\uparrow\uparrow}$ they read  $2\varepsilon = 2\omega_x \pm 2g\sqrt{N-1/2}$, being the energies of the second rung of the Tavis--Cummings ladder.
Finally, the poles of the amplitude $M_{\uparrow\downarrow}$ are determined by interaction of the second rung with the biexciton state $2\varepsilon = 2\omega_x-B$, see denominator of Eq.~\eqref{eq:Mpm}. Dependence of all the two-photon resonances  of the scattering amplitude on the biexciton binding energy is shown in Fig.~\ref{fig:ladder}. One can see that the poles of $M_{\uparrow\downarrow}$ (solid red lines) indeed originate from the anti-crossing of the poles of $M_{\uparrow\uparrow}$ (solid blue lines) and the biexciton state (dotted line).

{\it Photon blockade.}---We suppose that the system is excited with coherent $x$-polarized light of the frequency $\varepsilon$. If the light intensity is low the incoming state can be written in the form $|\psi_{\text{in}}\rangle =e^{\alpha a_{\varepsilon,x}^\dag } |0\rangle \approx (1+\alpha a_{\varepsilon,x}^\dag +\alpha^2 a_{\varepsilon,x}^{\dag 2}/2)|0\rangle$, where $a_{\varepsilon,x}^\dag$ is the creation operator for an $x$-polarized photon with the frequency $\varepsilon$ and $|\alpha|^2 \ll 1$ is proportional to the light intensity. Using the expression for the scattering matrix Eq.~\eqref{eq:S} we obtain the transmitted photon pair state
\begin{align}\label{eq:psi}
|\psi_{\text{out}}^{(2)}\rangle =  \frac1{2}t^2(\varepsilon)&\alpha^2 a_{\varepsilon,x}^{\dag 2} |0\rangle \\\nonumber
+\frac{\alpha^2}{4}\int \frac{d\omega}{2\pi} \big[ &M_{xx,xx}(\omega,2\varepsilon-\omega;\varepsilon,\varepsilon) a_{\omega,x}^\dag a_{2\varepsilon-\omega,x}^\dag  \\\nonumber
+&M_{yy,xx}(\omega,2\varepsilon-\omega;\varepsilon,\varepsilon) a_{\omega,y}^\dag a_{2\varepsilon-\omega,y}^\dag \big]
|0\rangle \,
\end{align}
and the transmitted one-photon state  
$
|\psi_{\text{out}}^{(1)}\rangle = t(\varepsilon)\alpha a_{\varepsilon,x}^\dag |0\rangle$,
where the scattering amplitudes 
in the basis of linear polarizations read $M_{xx,xx}= M_{\uparrow\downarrow} +\frac12 M_{\uparrow\uparrow}$, $M_{yy,xx}=M_{\uparrow\downarrow} -\frac12 M_{\uparrow\uparrow}$.
The
 output state Eq.~\eqref{eq:psi} contains polarization-entangled photon pairs being a superposition of two $x$-polarized photons and two $y$-polarized photons~\cite{Benson2000}. The polarization entanglement reflects  two coherent pathways of biexciton radiation~\cite{Langbein2013}, distinguishing the considered problem from  the polarization-independent scattering studied in Ref.~\cite{Fan2007}. Efficiency of two-photon blockade is characterized by the second-order correlation function of transmitted light. For $x$-polarized transmitted photons the function is defined as $g^{(2)}(t) = \langle \psi_{\text{out}^{(2)}}| a_{x}^{\dag}(0) a_{x}^{\dag}(t) a_{x}^{\vphantom{\dag}}(t) a_{x}^{\vphantom{\dag}}(0) |\psi^{(2)}_{\text{out}}\rangle / \langle \psi^{(1)}_{\text{out}}| a_{x}^\dag a_{x}^{\vphantom{\dag}} |\psi^{(1)}_{\text{out}}\rangle^2 $,
where $a_x(t) = \int a_{\omega,x} \e^{-\rmi\omega t} d\omega$. 
Under resonant excitation, the time dependence of the correlation function is rather obvious: it monotonously tends  to unity at the time scale of $1/(\Gamma_c+\Gamma_x)$.  So we focus on the value  at $t=0$.

\begin{figure*}[t]
 \includegraphics[width=.95\textwidth]{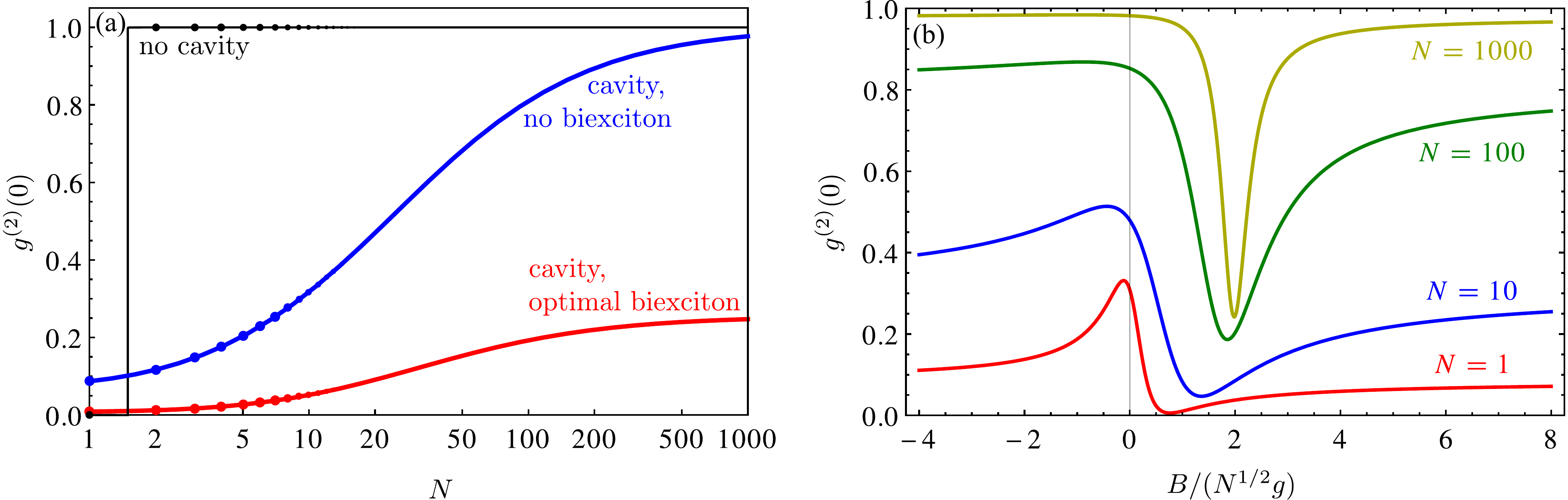}
 \caption{(a) Dependence of photon correlation function $g^{(2)}(0)$ on the number of QDs $N$. Thin black curve corresponds to reflection from QDs without cavity (or, alternatively, a cavity with QDs in a weak coupling regime). Thick blue and red curves show the correlation function of $x$-polarized transmitted photons for a cavity with $N$ QDs in the strong coupling regime, $\Gamma=0.1g$, under $x$-polarized excitation with the frequency $\varepsilon=\omega_x - \sqrt{N}g$. Thick blue curve corresponds to absence of biexciton state ($B=\infty$), while the thick red one corresponds to biexciton being tuned to the optimal energy, $B=B^*$, that maximizes the photon blockade. 
 (b) Dependence of $g^{(2)}(0)$  on the biexciton binding energy $B$ for cavity with different numbers of QDs in the strong coupling regime, $\Gamma=0.1g$. }\label{fig:g2B&N}
\end{figure*}

In what follows we consider the strong coupling regime $g\gg \Gamma_x,\Gamma_c$, since the weak coupling is equivalent to an absence of cavity.
We assume the excitation with the frequency $\varepsilon=\omega_x-\sqrt{N}g$, which corresponds to the one-particle resonance and maximizes one-photon transmission. This is in contrast to off-resonant excitation with the frequency $\omega_x$, considered in Ref.~\cite{Brecha1999}.
In our case 
\begin{align}
g^{(2)}(0) = \left| \frac{2N^{3/2}g-(2N+1)B+\frac{4\rmi N^2g}{g/\Gamma +2\rmi \sqrt N} }{4\rmi N^{3/2}g  - \sqrt{N}B(g/\Gamma +2\rmi \sqrt{N}) } \right|^2 \,, \label{eq:main}
\end{align}
where $\Gamma = \Gamma_c + \Gamma_x$. Expression Eq.~\eqref{eq:main} for the two-photon correlations is  our main result. Next, we analyze its dependence on  the number of dots $N$ (Fig.~\ref{fig:g2B&N}a) and the biexciton binding energy $B$ (Fig.~\ref{fig:g2B&N}b), aiming at the photon blockade condition $g^{(2)}(0)<1$.

The thick blue curve in Fig.~\ref{fig:g2B&N}a shows the behavior of $g^{(2)}(0)$ with $N$ for $B\to \infty$ corresponding to absence of biexciton resonance. At $N=1$ one has $g^{(2)}(0)\ll 1$ reflecting the photon blockade. 
The essence of the photon blockade effect is the detuning of the  two-photon transmission resonance from double energy of single-photon resonance~\cite{Kimble2005}.
Interestingly, the blockade is conserved even for multiple QDs in the cavity and destroyed 
only in the classical limit $N\to \infty$. This is in contrast to the case of QDs without a cavity where the blockade vanishes at $N\geq 2$, cf. thick blue and thin black curves in Fig.~\ref{fig:g2B&N}a. 
The photon pair in cavity could excite only the symmetric superradiant state 
of the Tavis--Cummings ladder $2\omega_x-2\sqrt{N-1/2}\,g$ (red curve in Fig.~\ref{fig:ladder}) that is however detuned from 
the double energy of excitation $2\varepsilon = 2(\omega_x - \sqrt{N}g)$ (yellow dashed line). 
The photon blockade  persists up to relatively large $N\sim (g/2\Gamma)^2\sim 25$ when  the detuning becomes less than the level width $\Gamma$. 

The effect of biexciton binding energy $B$ on the photon blockade  is shown in Fig.~\ref{fig:g2B&N}b. Surprisingly, the dependence of the second order correlation function on the biexciton energy is nonmonotonous: for certain binding energy $B^*$ the minimum of $g^{(2)}$ is achieved. Moreover, the dip in the dependence does not disappear with increase of $N$, meaning that the antibunching can be achieved even for large numbers of QDs. The reason is the biexciton-induced detuning of the two-particle superradiant state.
Namely, for $N \ll (g/2\Gamma)^2$ the photon blockade is enhanced due to the interference 
of the transmission channels of photons with parallel and opposite spins, $M_{xx,xx}= M_{\uparrow\downarrow} +\frac12 M_{\uparrow\uparrow}$.  The  maximal antibunching  $ g^{(2)}(0) = 
4(2N+1)^2 (\Gamma/g)^4 $ is achieved at the optimal binding energy 
$B^*=       N^{3/2}g/(N+1/2)$, see yellow dot in Fig.~\ref{fig:ladder}. This minimal value of $ g^{(2)}(0)$ is proportional to $(\Gamma/g)^4$ and parametrically smaller than that for the absence of biexciton resonance (at $B=\infty$ one has $g^{(2)}(0) = (2N+1)^2\Gamma^2/Ng^2$), cf. solid red and blue curves in Fig.~\ref{fig:g2B&N}a. 
For large $N\gg (g/2\Gamma)^2$ the correlation function $g^{(2)}(0)$ is equal to unity for all biexciton binding energies
except for the narrow region of the width $\sim g$ around $B^*= 2\sqrt{N} g$ corresponding to the anticrossing of the biexciton level with the second rung of the Tavis--Cummings ladder. In this case the double excitation energy hits exactly into the energy gap between the split $\uparrow\downarrow$ superradiant modes, see Fig.~\ref{fig:ladder}. The splitting $\sim g$ being larger than the level width $\sim \Gamma$,  the photon blockade is recovered in the $\uparrow\downarrow$ channel, i.e.
transmission of two counter-polarized photons is suppressed.
The $\uparrow\uparrow$ channel is not coupled to the biexciton resonance, and does not manifest the photon blockade. The total value of $g^{(2)}(0)$ for linearly polarized photons, contributed by both $\uparrow\uparrow$ and $\uparrow\downarrow$ channels, is equal to $1/4$, see the red curve in Fig.~\ref{fig:g2B&N}a.


To summarize, we have demonstrated that the photon blockade regime is realized for the polarization-dependent two-photon transport through a seemingly classical system of a microcavity with multiple quantum dots. Two-photon correlation function of transmitted photons is a nonmonotonous function of the  biexciton resonance energy. Even for a cavity with many quantum dots the proper tuning of biexciton energy allows one to achieve the values of $g^{(2)}(0)$ that are as low as $1/4$.   
Experimental realization of the blockade requires the inhomogeneous broadening of exciton frequencies to be small enough so that the collective superradiant exciton mode exists. This is feasible because  the simultaneous strong coupling regime of the cavity mode with up to three quantum dots has been already reported~\cite{kulakovskii2006,Laucht2010,Kim2011,Albert2013}.

\paragraph*{Acknowledgments.} 
The authors acknowledge fruitful discussions with M.~M.~Glazov and S.~A.~Tarasenko.
This work was supported by the RFBR,  
 and the Foundation ``Dynasty''.

\end{document}